\documentclass{aipproc}

\usepackage{graphicx} % Include figure files
\usepackage{dcolumn}  % Align table columns on decimal point
\usepackage{bm}       % bold math
\newcommand{\be}{\begin{equation}}
\newcommand{\ee}{\end{equation}}
\newcommand{\ba}{\begin{eqnarray}}
\newcommand{\ea}{\end{eqnarray}}
\newcommand{\bse}{\begin{subequations}}
\newcommand{\ese}{\end{subequations}}

\newcommand{\rvir}{r_{\textrm{\tiny{vir}}}}
\newcommand{\Mvir}{M_{\textrm{\tiny{vir}}}}

\layoutstyle{6x9}

\begin{document}
\title{Empirical testing of Tsallis' Thermodynamics as a model for dark matter 
halos
\footnote{To appear in the {\bf Proceedings of X Mexican Workshop on Particles 
and Fields}, Morelia Michoac\'an, M\'exico, November 7-12, 2005.}}


\author{Dario Nunez, Roberto A. Sussman, Jesus Zavala}
{address={Instituto de Ciencias Nucleares,\\
             Departamento de Gravitaci\'on y Campos,\\
             Universidad Nacional Aut\'onoma de M\'exico (ICN-UNAM).\\
             A. Postal 70-543, 04510 M\'exico, D.F., M\'exico.}}
\author{Luis G. Cabral-Rosetti}
{address={Departamento de Posgrado,\\
Centro Interdisciplinario de Investigaci\'on y 
Docencia en Educaci\'on T\'ecnica (CIIDET),\\
Av. Universidad 282 Pte., Col. Centro, A. Postal 752, C. P. 76000,\\
Santiago de Quer\'etaro, Qro., M\'exico.}}
\author{Tonatiuh Matos}{address={Departamento
de F{\'\i}sica, Centro de Investigaci\'on y de Estudios Avanzados
del IPN, \\ A.P. 14-740, 07000 M\'exico D.F., M\'exico.}}
\begin{abstract}
\noindent
We study a dark matter halo model from two points of view: the 
``stellar polytrope'' (SP) model coming from Tsallis' thermodynamics, and the
one coming from the Navarro-Frenk-White (NFW) paradigm. We make an appropriate
comparison between both halo models and analyzing the relations between the 
global physical parameters of observed 
galactic disks, coming from a sample of actual galaxies, with the ones of the 
unobserved dark matter halos, we conclude that the SP model is favored over 
the NFW model in such a comparison. 
\end{abstract}

\maketitle

\section{Introduction}

An alternative formalism to the micro-canonical ensemble treatment, that allows non-extensive
forms for entropy and energy under simplified assumptions, has been developed by Tsallis 
\cite{Tsallis} and applied to self--gravitating systems \cite{PL,TS1,TS2} 
under the assumption of a kinetic theory treatment and a mean field approximation.
As opposed to the Maxwell--Boltzmann distribution that follows as the equilibrium state 
associated with the  usual Boltzmann--Gibbs entropy functional, the Tsallis' functional
yields  as equilibrium state the ``stellar polytrope'' (SP), characterized by a polytropic 
equation of state with index $n$. 

On the other hand, high precision N--body  numerical simulations based
on Cold Dark Matter (CDM) models, perhaps the most powerful method available for 
understanding gravitational clustering, lead to the famous results of Navarro,
Frenk and White (NFW model) \cite{NFW} that predicts density and velocity profiles which are 
roughly consistent with observations, however, it also predicts a cuspy 
behavior at the center of galaxies that is not observed in most of the rotation
curves of dwarf and LSB galaxies \cite{Blok1,Blok2,Bin,B-S,B-O,Bosma3}. The 
significance of this discrepancy with observations is still under dispute, nevertheless, 
the NFW model of collision--less WIMPs remains as a viable model to account for DM
in galactic halos, provided there is a mechanism to explain the
discrepancies of this model with observations in the center of
galaxies.

Since gravity is a long--range interaction and virialized self--gravitating 
systems are characterized by non--extensive forms of entropy and energy, it is
reasonable to expect that the final configurations of halo
structure predicted by N--body simulations must be, somehow,
related with states of relaxation associated with non--extensive
formulations of Statistical Mechanics; therefore, a comparison between the SP
and NFW models is both, possible and interesting. The main purpose of our analysis 
is to make a dynamical analysis of
two halo models, one based on the NFW paradigm, and other based on
the SPs derived from Tsallis' non--extensive
thermodynamics, compare them and test both with observational
results coming from a sample of disk galaxies.

\section{SP and NFW halo models}

For a face space given by $({\bf{r}},\,{\bf{p}})$, the kinetic
theory entropy functional associated with Tsallis' formalism
is \cite{PL,TS1}, and \cite{TS2}
\begin{equation}
S_{q}\ =\ -\frac{1}{q-1}\,\int
{(f^{q}-f)\,{d}^{3}{\bf{r}}\,%
{d}^{3}{\bf{p}}},  \label{q_entropy}
\end{equation}
where $f$ is the distribution function and $q>1$ is a real number. In the
limit $q\rightarrow 1$, the functional (\ref{q_entropy}) leads to the usual
Boltzmann--Gibbs functional, corresponding to the isothermal sphere. 
The condition 
$\delta \,S_{q}=0$ leads to the distribution function that corresponds to the 
SP model characterized by the equation of state $p\ =\ K_n\,\rho ^{1+1/n}$,
where $K_n$ is a function of the polytropic index $n$, and can be expressed
in terms of the central parameters: $K_n=\frac{{\sigma_c}^2}{{\rho_c}^{1/n}}$.
The polytropic index, $n$, is related to the Tsallis' parameter $q>1$ by:
$n\ =\ \frac{3}{2}+\frac{1}{q-1}$. Inserting the equation of state into Poisson's
equation, the Lane-Emden equation \cite{B-T} is obtained, based on it we can obtain
density, mass and velocity profiles for the SP model.

NFW numerical simulations yield the following expression for the density
profile of virialized  galactic halo structures \cite{NFW,Mo}:
\begin{equation}
\rho_{_{\mathrm{NFW}}}=\frac{\delta_0\,\rho_0}{y\,\left(1+y
\right)^2},   \label{rho_NFW}
\end{equation}
where: $\delta_0=\frac{\Delta\,c_0^3}{3\left[\ln\,(1+c_0)-c_0/
(1+c_0)\right]}, \ \rho_0=\rho_{\mathrm{crit}}\Omega_0\,h^2 = 253.8 \, h^2\,\Omega_0\,
\frac{M_\odot}{\rm{kpc}^3}, \ y= c_0\frac{r}{r_v}$, and $\Omega_0$ is the ratio of 
the total density to the critical density of the Universe. Using equation
(\ref{rho_NFW}) its easy to obtain mass and velocity profiles.

\section{Comparison of SP and NFW models}

In order to compare both halo models, 
it is important to make various physically motivated assumptions.
First, we want both models to describe a halo of the same scale,
which means same virial mass $\Mvir$. Secondly, both models must have the same 
maximal value for the rotation velocity. This is a plausible assumption, as it is based
on the Tully--Fisher relation \cite{TF}, a very well established
result that has been tested successfully for galactic systems,
showing a strong correlation between the total luminosity of a
galaxy and its maximal rotation velocity. Our third assumption is
that the polytropic and NFW halos, complying with the
previous requirements, also have the same total energy evaluated at the
cut--off scale $r=\rvir$. The main
justification for this assumption follows from the fact that
the total energy is a fixed quantity in the collapse and subsequent
virialized equilibrium of dark matter halos~\cite{Padma1}. Since the SP model
has three free parameters (contrary to only one parameter of the NFW model), 
and we have selected three comparison criteria, we have a
mathematically closed problem once $\Mvir$ is specified.

\begin{figure}\centering
\includegraphics[height=6cm]{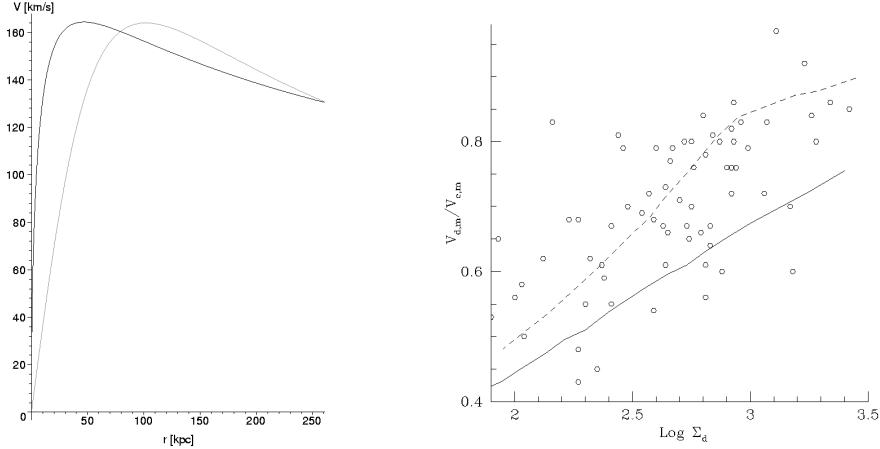}
\caption{Left panel: velocity profiles for the NFW halo (solid line) and its compared SP one
(dashed curve). Right panel: luminous to total dark matter content vs central surface density; 
open circles correspond to observational data, the dashed line represents the
NFW model and the solid one to SP model. Both models have a virial
mass of $M_v=10^{12}M_{\odot}$.}
\label{fig}
\end{figure}

Following the guidelines described above, we proceed to compare
NFW and polytropic halos for $\Mvir$ ranging from $10^{10}$ up to
$10^{15}$ solar masses. From this comparison we find the ``best-fit'' values 
for the free parameters of the SP model. The results are displayed explicitly 
in table \ref{tab:t1}. 
The comparison between both models in velocity profiles is shown in the left
panel of figure \ref{fig}.
SP and NFW models have both the same virial mass, $\Mvir = 10^{12}\,M_\odot$. 
For other values of $\Mvir$ the velocity profiles are qualitatively similar to
the one displayed in figure \ref{fig} (left panel). The detailed description and results of
the method of comparison presented above will appear in an article that is being prepared
\cite{Tsallis1}.

So far we have analyzed only the global structure of the dark matter halo
without considering the effects of the luminous galaxy within. If one wishes
to test a given model with observational results, it is necessary to add the
galactic baryonic disk as a dynamical component of the model. In order to do so we 
followed the method described in \cite{Mo, Tsallis3}. Then to compare both models with
observational results we used the prescription presented in \cite{JZ}.   

Using such a prescription, we may define the ratio of
maximum disk velocity to maximum total velocity,
$V_{d,m}/V_{c,m}$, which is a global quantity that can be directly
compared with theoretical predictions. This ratio is not defined at a given radius, but
it can be related to the total mass to disk mass ratio
$M_{t}/M_d$, defined at an specific radius. In particular at
radius $r_m$ where the total rotational curve has its maximum
we have \cite{JZ}:
\begin{equation}
\left(\frac{M_{t}}{M_d}\right)_{r_m}\propto
\left(\frac{V_{d,m}}{V_{c,m}}\right)^{-2}
\end{equation}
The use of $V_{d,m}/V_{c,m}$ instead of $M_{t}/M_d$ is
suitable because it  can be obtained directly
from observational parameters, without the assumptions needed to
calculate $M_{t}/M_d$.  

\begin{table}
\caption{Parameters characterizing the polytropes while being
compared to NFW halos}
\begin{tabular}{cccccccc}
$\log_{10}(\Mvir/M_{\odot})$ & $\rho_c\, [M_{\odot}/\rm{pc}^3]$ &
$\sigma_c\, [\rm{Km/s}]$  & $n$ & $q$ & $K_n$ & $v_{\rm{max}}\,
[\rm{Km/s}]$ & $\rvir\, [\rm{kpc}]$
\\ \hline 15 & $3.7 \times 10^{-4}$ & $982$ & $4.93$ & $1.29$ &
$4873.4$ & $1504$ & $2606.2$\\
12 & $7.5 \times 10^{-4}$ & $108$ & $4.87$ & $1.30$ &
$478.94$ & $164$ & $260.6$\\
11 & $9.0 \times 10^{-4}$ & $52$ & $4.83$ & $1.30$ &
$221.82$ & $79.1$ & $120.9$\\
10 & $1.2 \times 10^{-3}$ & $25$ & $4.82$ & $1.30$ &
$100.68$ & $38.2$ & $56.1$\\
\end{tabular}
\label{tab:t1}
\end{table}

One of the principal results obtained in the work \cite{JZ}
is that the ratio $V_{d,m}/V_{c,m}$ correlates principally with
the disk surface density $\Sigma_d$ of galaxies. Therefore we will use this result to 
compare the NFW and the SP models with the observational results coming from
the sample. We will take $\Mvir=1\times10^{12}M_{\odot}$, which is a characteristic value 
for the mass of dark matter halos. For simplicity, we will assume that the baryonic mass 
fraction of the disk, $f_d$, has the same value for all disks.

In the right panel of figure \ref{fig} it is shown the ratio 
$V_{d,m}/V_{c,m}$ vs $Log\Sigma_d$
for the sample of galaxies used (open circles). An almost
linear trend can be seen between this quantities. HSB (high
surface brightness) galaxies (corresponding approximately to
values of $Log\Sigma_d$ greater than 2.5) have greater values of
$V_{d,m}/V_{c,m}$ than LSB (low surface brightness) galaxies. This
means that the luminous matter content is greater for HSB than for
LSB galaxies. The shown picture is consistent with a well known
result: LSB galaxies are dark matter dominated systems within the
optical radius.

The value of the graphic presented is that it allow us to bound
statistically the possible values of the $V_{d,m}/V_{c,m}$ ratio
that galaxies with a given surface density can have. As was proved by 
\cite{JZ}, the size of this range of values (associated with dispersion
on the graphic) is mainly due different virial
mass that galaxies with the same $\Sigma_d$ have.

Right panel of figure \ref{fig} also show clearly that NFW models can not 
reproduce at satisfaction the results obtained for the compiled sample without
introducing unrealistic values for the virial mass. This is one of the results that lead us
to the possibility of seeking an alternative to the NFW paradigm.
The curves shown in figure \ref{fig} (right panel)
represent both models with average values for their respective
parameter; it's clear from the figure that the SP model follows in better 
agreement the average
behavior of the observational sample than the NFW model.

\section{Conclusions}

Motivated by the fact that SPs are the equilibrium state in
Tsallis' non--extensive entropy formalism, we have found the structural
parameters of those SPs that allows us to compare them with NFW
halos of virial masses in the range $10^{10}<\Mvir/M_\odot<10^{15}$; 
the results are displayed in Table \ref{tab:t1}. It is shown in the left panel
of figure \ref{fig} that the velocity profile of the SP model is much 
less steep in the same region
than that of the NFW halo. These features are consistent with the
fact that NFW profiles predict more dark matter mass
concentration than what is actually observed in a large sample of
galaxies~\cite{vera,Bin,JZ}.

We have also shown that the SP model
is favored over the NFW model regarding the dark matter content in disk 
galaxies (within the optical radius) which is shown by the average behavior of 
the observational sample in figure \ref{fig} (right panel). These results 
show that 
the NFW halo model can be enhanced with the
use of alternative paradigms in Statistical Mechanics, which seems to  
solve a recurrent item which throws a shadow in such an excellent description
as the NFW model is.

We are grateful to Vladimir Avila-Reese for his comments and suggestions 
to the manuscript of the present work. We acknowledge partial support by 
CONACyT M\'exico, under grants 32138-E 32138-E, 42748, 45713-F and 34407-E, 
and DGAPA-UNAM IN-122002 and IN117803 grants. TM wants to thank Matt Choptuik 
for his kind hospitality at the UBC, JZ acknowledges support from DGEP-UNAM 
and CONACyT scholarships.

%\section*{References}

%%%%%%%%%%%%%%%%%%%%%%%%%%%%%%%%%%%%%%%%%%%%%%%%%%%%%%%%%%%%%%%%%%%%%%%5

%\pagebreak

% Figuras y Tabla

%\begin{figure}
%\centering
%\includegraphics[height=7cm]{f2.eps}
%\caption{}
%\label{fig:2}
%\end{figure}

%%%%%%%%%%%%%%%%%%%%%%%%%%%%%%%%%%%%%%%%%%%%%%%%%%%%%%%%%%%%%%%%%%%%%%%%%%%%%%%%%%%%%%

%
\end{document}